\documentclass[aps,twocolumn]{revtex4}
\usepackage{graphicx}
\usepackage{tabularx}

\begin{document}

\title{Unambiguous Identification of the Second $2^+$ State in $^{12}$C and the Structure of the Hoyle State}

\author{W.R. Zimmerman$^{1,2}$, M.W. Ahmed$^{2,3}$, B. Bromberger$^4$, S.C. Stave$^2$, A. Breskin$^5$, V. Dangendorf$^4$, Th. Delbar$^6$, M. Gai$^{1,7}$, S.S. Henshaw$^2$, J.M. Mueller$^2$, C. Sun$^2$, K. Tittelmeier$^4$, H.R. Weller$^{1,2}$ and Y.K. Wu$^2$}
\affiliation{1. Dept of Physics, University of Connecticut, Storrs, CT 06269-3046 \\
2. Triangle Universities Nuclear Laboratory and  Dept of Phyaics, Duke University, Durham, NC 27708-0308 \\
3. Dept of Mathematics and Physics, North Carolina Central University, Durham, NC 27707 \\
4. Physikalisch-Technische Bundesanstalt, 38116 Braunschweig, Germany \\
5. Dept. of Particle Physics, Weizmann Institute of Science, 76100 Rehovot, Israel \\
6. Dept of Physics, Universite Catholique de Louvain, 1348 Louvain-la-Neuve, Belgium  \\
7. Dept of Physics, Yale University, New Haven, CT 06520-8124 }

\begin{abstract}
\      \\
The second J$^\pi \ = \ 2^+$ state of $^{12}$C, predicted over fifty years ago as an excitation of the Hoyle state, has been unambiguously identified using the $^{12}$C($\gamma,\alpha_0)^8$Be reaction. The alpha particles produced by the photodisintegration of $^{12}$C were detected using an Optical Time Projection Chamber (O-TPC). Data were collected at beam energies between 9.1 and 10.7 MeV using the intense nearly mono-energetic gamma-ray beams at the HI$\gamma$S facility. The measured angular distributions determine the cross section and the E1-E2 relative phases as a function of energy leading to an unambiguous identification of the second $2^+$ state in $^{12}$C at 10.03(11) MeV, with a total width of 800(130) keV and a ground state gamma-decay width of 60(10) meV; B(E2: $2^+_2 \ \rightarrow \  0^+_1) \ = \ 0.73(13)$ e$^2$fm$^4$ [or 0.45(8) W.u.]. The Hoyle state and its rotational $2^+$ state that are more extended than the ground state of $^{12}$C presents a challenge and constraints for models attempting to reveal the nature of three alpha particle states in $^{12}C$. Specifically it challenges the ab-initio Lattice Effective Field Theory (L-EFT) calculations that predict similar r.m.s. radii for the ground state and the Hoyle state.

\end{abstract}

\preprint{UConn-40870-00XX}

\maketitle

The second J$^\pi \ = \ 0^+$ state at 7.654 MeV in $^{12}$C, first predicted by Hoyle \cite{Hoyle} in 1953 and thus called the Hoyle state, plays a central role in nuclear physics. It is a well known fundamental testing ground of models of the clustering phenomena in light nuclei which is highlighted by recent developments of ab-initio theoretical calculations that are able to calculate light nuclei such as $^{12}$C. The Hoyle state plays a central role in stellar helium burning by enhancing the production of $^{12}$C in the universe allowing for life as we know it. It is the first and quite possibly still the best example of an application of the anthropic principle in physics. Early on after the discovery of the Hoyle state it was suggested by Morinaga \cite{Morinaga} that we can learn more about the structure of the Hoyle state by studying the rotational band built on top of it which led to a fifty year long  search for the second $2^+$ state in $^{12}$C \cite{Physics}. 

Recently the existence of the second $2^+$ state in $^{12}$C has been the subject of much debate. It was observed at approximately 9.8 MeV in measurements of the $^{12}$C$(\alpha,\alpha')$ and  the $^{12}$C(p,p') inelastic scattering reactions \cite{Osaka,iThemba,Yale,Freer}, but it was not observed below 10 MeV in either the beta-decays of $^{12}$N and $^{12}$B \cite{Beta} or in a recent  measurement of $^3$He induced reactions on $^{10,11}$B at 4.9 and 8.5 MeV \cite{Alc12}. In contrast, analysis of the beta-decay data \cite{Beta} suggest a $2^+$ state at 11.1 MeV which was not observed in the $^{12}$C(p,p') data \cite{Yale} or the recent measurement of the $^{11}$B($^3$He,d) reaction at 44 MeV \cite{Smit}. 

Previous measurements \cite{Osaka,iThemba,Freer,Beta} are dominated by the broad ($\Gamma \ \approx \ 3.0$ MeV) $0^+$ state at 10.3 MeV, as well as the narrow $3^-$ state at 9.641 MeV \cite{Osaka,iThemba,Yale,Freer,Beta}. Indeed a $2^+$ state below 10.0 MeV was observed in the inelastic scattering data \cite{Osaka,iThemba} only after separating from a large background contribution from the third $0^+$ at 10.3 MeV. Gamma-ray beams as used in this study \cite{HIgS} cannot populate $0^+$ states and will populate the $3^-$ state with very small probability making them an excellent probe to use in the search for $2^+$ state in $^{12}$C. Since $\gamma$-ray beams can also induce E1 transitions the nearby $1^-$ state at 10.84 MeV is expected to contribute. In the current study we used the Optical Time Projection Chamber (O-TPC) detector discussed in \cite{JINST} to detect the outgoing particles with nearly 100\% efficiency. Thus our gamma-ray beam plus an O-TPC detector system is a powerful (beam-target-detector) combination in the search for $2^+$ states in $^{12}$C. In this paper we present background free data with an unambiguous identification of the second $2^+$ state at 10.03(11) MeV in $^{12}$C.

Ever since Brink suggested that the Hoyle state is a very extended object with the structure of three alpha-particles arranged in a linear chain \cite{Brink} many  theoretical models have been developed to describe the Hoyle state and the structure of $^{12}$C. One of the issues of great current interest is the r.m.s. radius of the Hoyle state and in particular whether the Hoyle state is an extended object with an rms radius considerably larger than the ground state of $^{12}$C.

A number of models have been proposed to describe the structure of $^{12}$C including an algebraic U(7) model with a D$_{3h}$ symmetry \cite{U7}, a microscopic Fermionic Molecular Dynamic (FMD) model \cite{FMD} together with a "BEC-like" cluster model \cite{BEC} which predicts wave functions that are quite similar to those previously predicted by the RGM cluster model \cite{Kamim}, an ab-initio No-Core Shell Model (NCSM) \cite{abinitio}, and a No-Core Simplectic Model (NCSpM) \cite{Draayer}, and ab-initio Lattice Effective Field Theory calculations (L-EFT) \cite{EFT,Dean}. The NCSM calculations that currently extend up to 10 $\hbar \omega$ do not adequately predict the location of the Hoyle state \cite{abinitio} and in the symmetry inspired schematic NCSpM calculations \cite{Draayer} a model space of up to 20 $\hbar \omega$ is needed to predict the low lying Hoyle state close to the measured energy. The many $\hbar \omega$ model space required in these shell model calculations is suggestive of a very extended alpha-clustering configuration. Such an extended state arises naturally in cluster models \cite{U7,FMD,BEC,Kamim} and it represents a major challenge to ab-initio calculations \cite{abinitio,EFT,Dean}.

Current models differ on the shape of the Hoyle state. In the U(7) model \cite{U7} and the FMD model \cite{FMD} the Hoyle state is predicted to be an oblate equi-lateral triangular three alpha-particle configuration. Both U(7) and FMD models predict a rotational band built on the Hoyle state but the FMD model predicts the second $2^+$ in $^{12}$C to have a B(E2) to the third $0^+$ state which is twice as large as that leading to the Hoyle state, and thus not to be a member of the Hoyle rotational band. In the L-EFT calculations \cite{Dean} the Hoyle state is primarily of the bent-arm chain (or obtuse triangular) shape. Both NCSpM calculations  \cite{Draayer} and L-EFT calculations \cite{Dean} predict the Hoyle state to have a deformed prolate shape with a rotational band built on it. The "BEC-Like" cluster model \cite{BEC} predicts the Hoyle state to be spherically symmetric. In addition while the FMD model predicts an rms radius of the Hoyle state (3.38 fm) that is $\sqrt 2$  larger than the ground state (2.34 fm) \cite{FMD}, the ab-initio L-EFT calculations \cite{Dean} predicts an rms radius (2.4 fm) equal (within the predicted error bar) to the rms radius of the ground state of $^{12}$C. The NCSpM calculations \cite{Draayer} predict an rms radius (2.93 fm) that is 25\% larger than the ground state of $^{12}$C.

The three alpha-particle structure of $^{12}$C naturally leads to models that utilize triangular geometry \cite{U7,FMD,EFT,Dean}. Such triangular systems are ubiquitous in physics including the X$_3$ molecular system \cite{X3} and the three quark system \cite{QCD1,QCD2}, and their spectra resemble the one predicted by the oblate spinning top with a D$_{3h}$ symmetry \cite{X3,QCD1}. Unlike molecules, in nuclei the energy scale of rotations and vibrations are similar, leading to large mixing of rotational and vibrational states which leads to deviation from the prediction of a rigid rotor \cite{U7}. Still the phenomenological schematic U(7) model preserves the rotation-vibration structure \cite{U7} and it serves as a useful guiding tool  for discussing the essential degrees of freedom of the three alpha-particle system.

The current measurement of the $^{12}$C($\gamma,\alpha)^8$Be reaction was performed at the HI$\gamma$S facility that produces an intense, nearly monoenergetic gamma-ray beam by Compton backscattering photons of free-electron laser \cite{HIgS}. Beams of circularly polarized gamma-rays with energies between 9.1 and 10.7 MeV were used with energy spreads of 300 - 350 keV and on-target intensities of  $\approx 2 \times 10^8 \ \gamma /sec$. The beam intensity was measured by detecting neutrons from the d$(\gamma$,n)p reaction using an in-beam D$_2$O target, cross-calibrated against a large NaI(Tl) detector. The energy profile of the beam was measured using a large HPGe detector, and the spectra were unfolded using a Monte Carlo technique \cite{Sun1,Sun2}. The alignment of the detector with respect to the beam was achieved using a gamma camera and lead absorbers placed in the front and back of the detector as discussed in \cite{JINST}. 

\begin{figure}
 \includegraphics[width=3in]{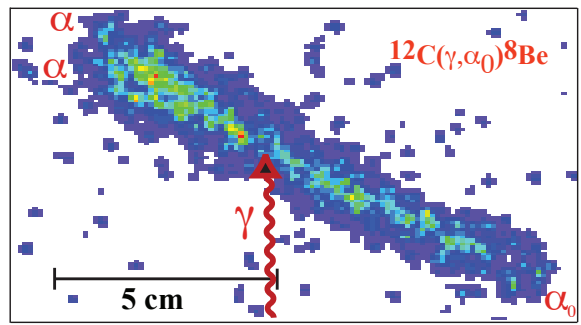}
 \caption{\label{Track} (Color Online) A typical image recorded by the CCD camera of three alpha-partciles from the reaction $^{12}$C$(\gamma,\alpha_0)^8$Be $(\rightarrow \ \alpha \ + \ \alpha)$.}
\end{figure}

An O-TPC operating at 100 torr with the gas mixture of CO$_2(80\%) \ + $ N$_2(20\%)$  \cite{JINST} was used to detect the outgoing alpha-particles from the $^{12}$C$(\gamma,\alpha)^8$Be reaction. The events recorded in the O-TPC include protons from the $^{14}$N$(\gamma,p)$ reaction, alpha particles from the $^{16,18}$O$(\gamma,\alpha)$ and the $^{12}$C$(\gamma,\alpha)^8$Be reactions, and cosmic rays.  Nearly all (98\%) of the $^{12}$C dissociation events were observed to proceed via the $^{12}$C$(\gamma,\alpha_0)^8$Be reaction leading to the ground state of $^8$Be, and the subsequent immediate decay to two nearly co-linear alpha-particles as shown in Fig. 1. The $^{12}$C$(\gamma,\alpha)^8$Be events were easily separated from the other events listed above except for the  $^{16}$O$(\gamma,\alpha)$ events, using the energy deposited in the detector (with a measured energy resolution of FWHM $\approx$ 100 keV \cite{JINST}). The recorded energy and track from the $^{12}$C and $^{16}$O dissociation events are very similar but the events can be distinguished using the line-shape analysis of the time projection signals recorded by the photomultiplier tubes \cite{JINST}. 

Each measured (PMT) time projection signal was fit using the calculated line shapes of the $^{12}$C$(\gamma,\alpha)^8$Be and $^{16}$O$(\gamma,\alpha)^{12}$C events \cite{JINST}.  A good fit of all $^{12}$C and $^{16}$O events was obtained. The goodness-of-fit parameters, $\chi ^2_C$ and $\chi^2_O$ of the predicted line shapes of $^{12}$C$(\gamma,\alpha)^8$Be and $^{16}$O$(\gamma,\alpha)^{12}$C respectively, were used to classify the events as shown in Fig. 2 for the beam energy of 9.8 MeV. All events to the left of the dotted (red) line were identified as $^{12}$C$(\gamma,\alpha)^8$Be events. The efficiency of the cut and leakage of $^{16}$O$(\gamma,\alpha)^{12}$C events were estimated by fitting the distribution to the sum (shown by solid black line) of two log-normal functions (shown by dashed blue lines). The cut was placed such that fewer than 0.5\% of the events to the left of the cut were $^{16}$O$(\gamma,\alpha)^{12}$C events.

\begin{figure}
 \includegraphics[width=3in]{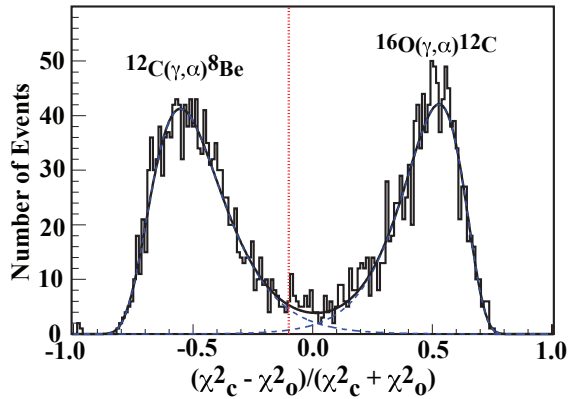}
 \caption{\label{EventID} (Color Online) Event identification function (E$_\gamma$ = 9.8 MeV) derived from the goodness-of-fit parameters ($\chi^2_C$ and $\chi^2_O$ for $^{12}$C and $^{16}$O dissociation events, respectively) as discussed in the text.}
\end{figure}

Complete angular distributions of $^{12}$C$(\gamma,\alpha_0)^8$Be events were measured at seven energies between 9.1 and 10.7 MeV. The events recorded in the O-TPC are transformed to the $(\theta,\phi)$ coordinate system [12] with an accuracy in $\theta$ varying between $2.5^\circ$ and $6.0^\circ$, depending on the out-of-plane angle of the track. The angular distributions were fit in terms of E1 + E2 amplitudes and their relative phase $\phi_{12}$ as discussed in section 4.1 of  \cite{Dyer}. Since angular information was available for each $^{12}$C$(\gamma,\alpha_0)^8$Be event individually, unbinned maximum likelihood fits were used to avoid losing information through binning. Angular distributions measured at gamma-ray beam energies of of 9.6 and 10.7  MeV are shown in Fig. 3  along with the fit yielding the cross section ratio of ${\sigma (E2) \over \sigma} \ = \  0.97^{+0.01}_{-0.02}$ and $0.71^{+0.04}_{-0.05}$, as well as phase angles $\phi_{12} = 80 \pm 6^\circ$ and $132 \pm 5^\circ$, respectively. The angular distributions were dominated by the E2 component at all but the highest beam energy (at 10.7 MeV) where a non-negligible contribution of the $1^-$ state at 10.84 MeV leads to a very asymmetric angular distribution, as shown in Fig. 3.

\begin{figure}
 \includegraphics[width=3in]{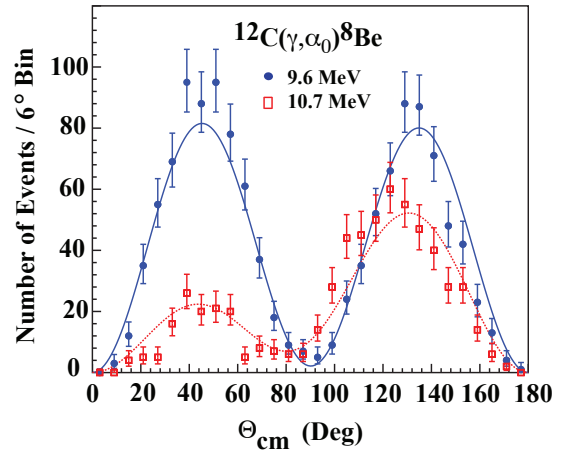}
 \caption{\label{AngDis} (Color online) Angular distribution for $^{12}$C$(\gamma,\alpha_0)^8$Be events measured at a beam energy of 9.6 and 10.7 MeV. The solid curve is the fit that included E1 and E2 amplitudes as discussed in the text. The error bars are statistical only.}
\end{figure}

The total E1 and E2 cross sections and the relative E1-E2 phase ($\phi_{12}$) extracted using the angular distribution data (as shown in Fig. 3) are shown in Fig. 4 as a function of energy with error bars that include both statistical and systematic uncertainties. The systematic uncertainties associated with each measured cross section are dominated by a 5\% uncertainty in the gamma-ray beam intensity. In Fig. 4(a) we show the E1 and E2 cross section components measured at these energies together with fits to Breit-Wigner resonances with energy-dependent level shifts and widths \cite{Lane}, convoluted with the measured gamma-ray beam energy distribution. Coulomb wave functions were calculated using the continued-fraction expansion technique \cite{Coul} with $r_0 \ = \ 1.4$ fm. The fit to the E1 cross section data uses the previously determined energy and width of the $1^-$ state at 10.84 MeV in $^{12}$C \cite{Ajz}, with the strength adjusted to fit the data. The fit to the E2 data includes three free parameters: the partial widths ($\Gamma_\alpha, \ \Gamma_\gamma$) and the resonance energy. 

\begin{figure}
 \includegraphics[width=3in]{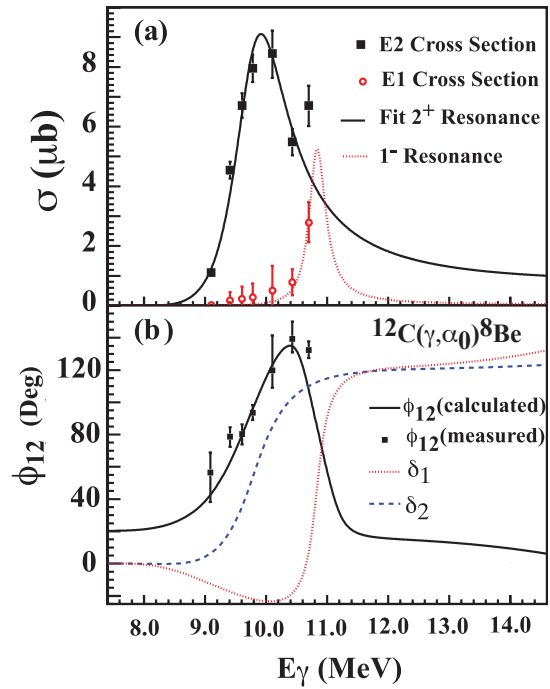}
 \caption{\label{Exc} (Color online) (a) The measured E1 and E2 cross sections of the $^{12}$C$(\gamma,\alpha_0)^8$Be reaction. (b) The measured E1-E2 relative phase angle ($\phi_{12}$) together with the phase angle calculated from a two-resonance model.}
\end{figure}

The E2 cross section data allow us to identify a $2^+$ resonance at 10.03(11) MeV with a total (alpha-particle) width of 800(130) keV (that exhausts 65(9)\% of the Wigner limit) in agreement with Ref \cite{Freer}. The measured gamma-decay width to the ground state is 60(10) meV leading to a reduced quadrupole electromagnetic transition of $B(E2: \ 2^+_2 \rightarrow 0^+_1) \ = \ 0.73(13)$ e$^2$fm$^4$ [or 0.45(8) W.u.]. This measured B(E2) is not too different from the prediction of the FMD model (0.46 e$^2$fm$^4$) \cite{FMD}, but somewhat smaller than predicted in the L-EFT calculations [2(1) e$^2$fm$^4$] \cite{Dean}.  Note that the slight difference between the maximum of the calculated cross section (at 9.8 MeV) and the resonance energy (at 10.03 MeV) is due to the energy-dependent widths used in the fit which push the maximum yield toward lower energies. The highest energy data point at 10.7 MeV seems inconsistent with this single resonance. In order to estimate the error in the measured resonance energy, we also analyzed our data including another $2^+$ that was previously suggested at 11.2 MeV \cite{Beta} leading to a total error in the resonance energy of 110 keV. The current results do not allow us to place any constraints on $2^+$ states at energies above 11 MeV.

The E1-E2 phase differences ($\phi_{12}$) extracted from our measured angular distributions are shown in Fig. 4(b). The measured phase angle is compared to the predicted phases \cite{Dyer}:
\begin{center}
$\phi_{12} \ = \ \delta_2 \ - \ \delta_1 \ + \ arctan(\eta /2)$
\end{center}
where the nuclear phase shifts $\delta_\ell$ are given by the resonance phase shift minus the hard sphere phase shift \cite{Lane} and $\eta$ is the Sommerfeld parameter. The $\ell$ = 1 resonance phase shifts ($\delta_1$) were calculated using parameters of the known $1^-$ state at 10.84 MeV \cite{Ajz} and the dip is due to the hard sphere contribution. The $\ell$ = 2 phase shifts ($\delta_2$) were calculated using the resonance energy and width determined from the fit to the E2 cross section data. The calculated $\phi_{12}$ curve was averaged over the measured cross section and gamma-ray beam energy distribution. The good agreement between the measured and calculated phases unambiguously establishes the existence, the energy and the width of the $2^+$ resonance reported here and indicates that there is little or no contribution from other (background) amplitudes.

The measured second $2^+$ state at 10.03(11) MeV reported in this work lies 2.38(11) MeV above the $0^+$ Hoyle state which is approximately half the excitation energy of the first $2^+$ state of $^{12}$C. Since the U(7) model predicts the ground state rotational band  and the Hoyle bands to arise from the same geometrical shape we conclude that in this model the radius parameter of the Hoyle state is approximately $\sqrt 2$ larger than the r.m.s. radius of the ground state. This conclusion is  consistent with (but slightly larger than) the r.m.s. radius of the Hoyle state [2.89(4) fm] determined from inelastic light-ion scattering experiments of  \cite{Dan09}. It should be noted that the present result and all previous determinations of the rms radius of the Hoyle state using $^{12}$C(e,e') data \cite{FMD} and $^{12}$C(x,x') data \cite{Dan09} are model dependent. Our inability to measure elastic scattering off the (very short lived) Hoyle state makes a direct measurement of the r.m.s. radius of the Hoyle state unlikely.

In conclusion we have used an Optical Time Projection Chamber (O-TPC) with intense nearly mono-energetic gamma-ray beams to unambiguously identify the long sought after second $2^+$ state in $^{12}$C having an extended three alpha-particle configuration. We provide the resonance parameters including the spin and parity ($J^\pi = 2^+$), energy, total (alpha-particle) width, and the B(E2) value for the decay to the ground state which must be reproduced by theoretical models which properly describe the low lying three alpha-particle structure of $^{12}$C.

\vspace{-0.5cm}

\section{Acknowledgement}


The authors would particularly like to thank the staff of HI$\gamma$S at Triangle Universities Nuclear laboratory (TUNL) for the operation of the facility, as well as T. Neff and D. Lee  for helpful discussions. One of us (MG) wishes to acknowledge discussions with F. Iachello. This work is supported in part by the U.S. Department of Energy, grant numbers DE-FG02-97ER41033 and DE-FG02-94ER40870.

\end{document}